# *In-situ* electronic characterization of graphene nanoconstrictions fabricated in a transmission electron microscope

Ye Lu[†], Christopher A. Merchant[†], Marija Drndić, A. T. Charlie Johnson

Department of Physics and Astronomy, University of Pennsylvania, Philadelphia, Pennsylvania, 19104.

[†]These authors contributed equally to this work.

Correspondence to Marija Drndić email: drndic@physics.upenn.edu
Correspondence to A. T. Charlie Johnson email: cjohnson@physics.upenn.edu

We report electronic measurements on high-quality graphene nanoconstrictions (GNCs) fabricated in a transmission electron microscope (TEM), and the first measurements on GNC conductance with an accurate measurement of constriction width down to 1 nm. To create the GNCs, freely-suspended graphene ribbons were fabricated using few-layer graphene grown by chemical vapor deposition. The ribbons were loaded into the TEM, and a current-annealing procedure was used to clean the material and improve its electronic characteristics. The TEM beam was then used to sculpt GNCs to a series of desired widths in the range 1 – 700 nm; after each sculpting step, the sample was imaged by TEM and its electronic properties measured *in-situ*. GNC conductance was found to be remarkably high, comparable to that of exfoliated graphene samples of similar size. The GNC conductance varied with width approximately as $G(w) \simeq (e^2/h)w^{0.75}$, where $w$ is the constriction width in nanometers. GNCs support current densities greater than 120 μA/nm$^2$, two orders of magnitude higher than has been previously reported for graphene nanoribbons and 2000 times higher than copper.

Graphene [1,2] is a unique material that may soon find use in electronic [3-5], biological [6-8], and optoelectronic [4,9,10] applications. However, the production of large-scale, high quality graphene for electronic devices remains an important challenge. To date, high mobility graphene devices have been primarily fabricated using mechanically-exfoliated graphene flakes [11,12], a technique that requires time-intensive characterization procedures [13] and is not scalable. Large-scale graphene production methods, such as chemical vapor deposition (CVD) on polycrystalline Cu foil, may be used to create wafer-scale graphene [14], however the reported mobility of the material to date is at least one order of magnitude lower than that of exfoliated graphene [14-17]. Methods for production of large-area graphene with high mobility are needed in order for electronic applications of graphene to be realized.

Bulk graphene also lacks an energy band gap, making it unsuitable for digital logic applications. A band gap can be engineered into the material by fashioning it into nanoribbons [18-20], nanoconstrictions [21], and quantum dots [22], which are predicted to develop energy gaps due to quantum confinement. The value of the energy gap is modified by edge effects [23-25] that are predicted to depend on atomic-scale details of the edge geometry. This motivates the development of techniques for measuring the electronic properties of graphene nanostructures whose structure may also be imaged with sub-nanometer resolution.

To address these issues, we developed the use of electron beam "nanosculpting" to fabricate graphene nanoconstrictions (GNCs) inside a TEM and measure their widths and electronic properties *in situ*. Compared to earlier reports based on scanning electron microscopy [21] and atomic force microscopy [18,26], our TEM-based approach enables superior control and accurate measurement of the GNC width across a very wide range (1 – 1000 nm). Current annealing is used to clean the graphene and improve its electronic properties. Based on measurements of the resistance of multiple GNCs sculpted from a single nanoribbon, we infer both the contact resistance and the intrinsic conductance of the GNC as a function of its width, information that was not accessible in the earlier experiments. We find that the intrinsic GNC conductance varies with the channel width $w$ approximately as $w^{0.75}$ over the range 1 – 700 nm, indicating that the GNCs experience carrier doping so the chemical potential is well away from the charge neutrality point. Current-annealed GNCs have high conductance that is comparable to that of exfoliated graphene samples with similar dimensions, and they withstand a current density of at least 120μA/nm$^2$, two orders of magnitude larger than has been reported previously on graphene nanoribbons [26,27]. This work thus provides more understanding of the correlation between graphene structure and performance, and it opens a route towards the study of the effect of atomic-scale edge geometry on the transport properties of graphene nanostructures.

The experiments were based on suspended graphene ribbons on electron-transparent SiN membranes, shown schematically in Figure 1 (a). Device fabrication began by micromachining a 40×40μm$^2$, 100 nm thick, freely-suspended SiN membrane, supported by a bulk silicon wafer [28]. A combination of photolithography and electron beam lithography was used to define Au source and drain electrodes with separations of 200 nm – 1 μm, followed by Focused Ion Beam milling to cut a 1.4 μm x 200 nm slit in the SiN membrane in the region between the electrodes. CVD-grown few layer graphene (FLG), ~3-10 layers thick, was transferred to the chip surface [17,29] and patterned using electron-beam lithography on negative-tone resist (XR-1541, Dow Corning) and an oxygen plasma etch. This resulted in FLG nanoribbons (typically 3 μm long by 400-800 nm wide) with a small freely-suspended region over a slit in the SiN membrane, in good electrical contact with the Au electrodes. The negative-tone resist was removed with a buffered oxide etch immediately before inserting the chip into the TEM on a home-built sample holder with electrical feedthroughs.

Bright field imaging and electrical measurements were performed at room temperature in a JEOL 2010F TEM. An image of a typical suspended FLG nanoribbon is shown in Fig. 1 (b). Two arrows indicate the edges of the material, which is quite faint in this image because it is only a few layers thick. The current-voltage (I-V) characteristic, measured *in-situ*, is shown in the inset of Fig. 1 (a). The initial resistance values of the nanoribbons used in these experiments were in the range 20 – 200 kΩ.

A current annealing (Joule heating) procedure with a slow voltage ramp (~ 10 mV/s) was performed on the nanoribbon inside the TEM; a typical I-V characteristic for the annealing process is shown in Fig. 2 (a), where red arrows indicate the sweep direction. The Joule power as a function of time is given in the inset of Fig. 2 (a). The maximum dissipated power exceeds 1.2 mW, which is expected to cause the suspended nanoribbon to reach a temperature as high as 2000 °C [30]; the high vacuum environment of the TEM chamber (~10 µTorr) prevents oxidation of the FLG [30,31]. We observe that the I-V is smooth for *V*< 2 V, with pronounced current jumps occurring at higher bias (i.e., a small and large current jump at 2.3V and 2.7V, respectively, in Fig. 2 (a)) [30]. After the annealing process, the I-V characteristic is permanently altered, as evidenced by the hysteresis in Fig. 2 (a).

The evolution of the two-terminal resistance (R=V/I) for the device during current annealing is plotted in Fig. 2 (b). Annealing results in a decrease in device resistance by more than an order of magnitude, from ~150 kΩ to ~10 kΩ. This is evident in the data of Fig. 2 (b) (taken at variable, relatively high bias voltage), as well as the low-voltage I-V characteristics measured before and after annealing (inset to Fig. 2(b)).

We attribute the current jumps and significant decrease in resistance that occur during current annealing to three factors. First, Joule heating of the sample to ~300-400°C is expected to cause vaporization of resist residue from the lithography processes and an associated increase in carrier mobility [32,33]. This temperature range is consistent with an applied voltage of 2.3V, where the smaller current jump is observed in Fig. 2(a). TEM images of FLG nanoribbons taken before and after annealing clearly indicate contamination removal as impurities evaporate from the surface (see Fig 1S in the Supporting Information). Second, high-temperature annealing has been shown to reduce the contact resistance between metal electrodes and carbon nanomaterials [34]. Finally, we observe that high-temperature current annealing induces a structural reconfiguration and recrystallization of the FLG ribbon, as observed by others [30,31] and discussed in the online Supporting Information. Current annealing thus significantly improves the structural and electronic properties of the CVD graphene.

After current annealing, nanosculpting with the focused electron beam [35] was used to define nanoconstrictions (GNCs) from the suspended ribbon. Beam irradiation was used to knock out carbon atoms from the ribbon edges and in this way gradually reduce the ribbon width. In order to nanosculpt the ribbon, the TEM magnification was increased to ~ 800,000x, the electron beam was focused to its minimum diameter, ~ 1nm, and the beam was moved with the condenser deflectors to expose and remove graphene at a rate of ~ 1 nm$^2$/s.

Nanosculpting was performed with an applied electrical bias voltage to heat the sample, which suppresses deposition of amorphous carbon [31] and damage of the lattice by the electron beam [36,37]. Figure 3(a) shows I-V traces taken during a set of four nanosculpting steps, where a nanoribbon was progressively narrowed from 280 nm to 14 nm to yield GNCs of different widths. To begin nanosculpting, the bias voltage was ramped to ~ 2.3 V and the current allowed to stabilize. The focused TEM beam was used to remove graphene from the sample edge, leading to a current decrease (vertical arrow). After each sculpting step, the low-bias I-V characteristic was measured with the TEM beam off to prevent carbon deposition (inset to Fig. 3(a)) and the GNC was imaged (Fig. 3(c) – (f)). Data correlating GNC width and electrical resistance from multiple samples is analyzed below.

The width of the sample in Fig. 3 was reduced to 5 nm by a final nanosculpting step, and the device was allowed to break under the stress of the applied voltage (Fig. 3(g) – (h)). Figure 3 (b) shows the time evolution of the device conductance from a value of ~ 3 $e^2/h$ to zero when the nanoconstriction breaks (the contact resistance has been subtracted off, as described below). The inset to Fig. 3(b) is a higher resolution view of how the device conductance evolves during the last few seconds before the constriction fails. In contrast to the breaking of metallic junctions, where atom-by-atom removal leads to quantized steps in the conductance [38], GNCs break in a less controlled way, consistent with the existence of strong covalent bonds between the carbon atoms. Junctions often break when they are several carbon atoms wide (e.g., Fig 3 (g) – (h)), and the conductance value immediately before the break shows significant device-to-device variation.

The TEM nanosculpting procedure enabled reliable fabrication and inspection of GNCs with arbitrary widths as small as 1 nm. Electronic measurements taken during and after the sculpting procedure provided correlated data on GNC conductance and current density as a function of GNC width. For the device highlighted in Fig (3), we estimate the GNC thickness of 2 layers (0.6 nm) and a final width of 5 ± 0.5 nm at its narrowest point, implying that this particular GNC supported a current density in excess of 30 µA/nm$^2$. The other GNC samples measured during this work supported current densities of 20 – 120 µA/nm$^2$ without failure, two orders of magnitude higher than has been previously reported for graphene nanoribbons [27,39] and ~2000 times higher than copper.

We attribute the high current density to several factors. First, we expect that current annealing leads to improved quality of the FLG material. This assumption is supported by the observation that GNC conductance is comparable to that found for samples we created by mechanical exfoliation. For example, we measure a conductance of 500 μS (1300 μS) for a 20-nm wide (50-nm wide) GNC, and we use data from an earlier report [19] and find a value of 750 μS (2000 μS) for exfoliated samples of similar length and width. Second, we assume that our GNCs are shorter than the mean free path for phonon scattering and therefore support quasi-ballistic electron transport, with most of the dissipation occurring in the bulk leads rather than the GNC; this phenomenon will lead to enhanced current carrying capacity, as is known to occur for carbon nanotubes [40]. Finally, based on Fig. 3 (g) and similar images (see Supporting Information), we expect that current annealed GNCs have crystallographically oriented edges, which sharply reduces edge scattering and enhances conduction. It is also possible that stable nanotube structures form in the narrowest GNCs [31], providing an additional conduction channel and increased current density. Future work with an aberration-corrected TEM will provide more definitive structural information on the GNCs and enable comparison of the measurements with experimental[41] and computational[42] work on graphene quantum point contacts.

Figure 4(a) shows two-terminal resistance, $R_{TOT}$, as a function of width for four sets of GNCs, each fabricated from a suspended nanoribbon samples using the TEM nanosculpting process outlined above. For each GNC set, we fit the data with the form $R_{TOT} = R_C + R_M w^{-\alpha}$, where $w$ is the measured width of the GNC (in nanometers), and the contact resistance, $R_C$, $R_M$ (with units of resistance) and $\alpha$ are fitting parameters. The contact resistance is expected to include contributions from the wiring, the gold-graphene contact interface, and wide graphene regions outside of the GNC. Although a natural expectation for the power law parameter is $\alpha = 1$, appropriate for an ohmic conductor, this value provides a consistently poor fit to the data, with unphysical values for the contact resistance (see Supporting Information for details). We find far superior fits to the data for $\alpha = 0.75$, as illustrated in Fig. 4(a).

Figure 4(b) shows conductance versus width for all the GNC data, after removal of the best-fit contact resistance, $R_C$. The full data set is well fit over its entire range by the functional form $G = \sigma_0 w^{0.75}$, where $\sigma_0 = e^2/h$, and $w$ is GNC width measured in nanometers. Choices of the power law parameter outside the range $\alpha = 0.7 - 0.8$ lead to fits of significantly lower quality (see Supporting Information).

This is a surprising observation in several ways. First, as mentioned earlier, we expect the confinement to induce an energy gap in the material, but there is little evidence of this in the data, where very high conductance is observed for the GNC down to a width of 1 nm. This observation is consistent with the presence of unintentional carrier doping such that the chemical potential for the system lies outside the energy gap. It is notable that the confinement-induced energy gap of a bare (unterminated) graphene nanoribbon is computed to vary as $E_g \sim w^{-b}$, with the power law $b$ in the range $0.75 - 0.9$ [24,25]. To reconcile this scaling with our finding that $G \sim w^{0.75}$, we would need to assume that $E_g \sim 1/w$. However, we are not aware of an established relationship between the energy gap and the conductance at large Fermi energy. Second, contrast changes observed in TEM images of GNCs (Figs. 3 (c) - (h), and especially 3 (f)), indicate the material becomes thinner as it is narrowed by nanosculpting. Based on the quality of the power law fit even as the GNC thickness varies, we conclude that the bulk of the conduction is carried by only a few (1 or 2) of the graphene layers that comprise the GNC, which are recrystallized by the annealing process (see Supporting Information), and perhaps thermally bonded to the Au electrodes. The TEM fabrication and measurement capability introduced here will enable future experiments to clarify these issues.

In summary, we have presented a robust methodology for the fabrication and *in-situ* measurement of graphene nanoconstrictions inside a TEM. GNCs formed by nanosculpting have high conductance comparable to that of exfoliated graphene, and they are able to sustain current densities in excess of 100 μA/nm$^2$. We measured the electrical properties of GNCs with widths in the range 1 – 1000 nm, and found that the GNC conductance varies approximately as *$w^{0.75}$*. Future work, including the use of an aberration-corrected TEM to allow for single-atom resolution imaging, may further illuminate the properties of GNCs and enable correlation of the sample conductance with the precise edge structure.

**Acknowledgment.** The authors thank Dr. Zhengtang Luo for useful discussions and development of the CVD graphene growth process used in these experiments; and Drs. Julio Rodriguez-Manzo and Branislav Nikolic for comments on the manuscript. This work was supported by the National Science Foundation through grant #DMR-0805136 (YL and ATCJ) and NIH Grant R21HG004767 (CM and MD). Use of facilities of the Nano/Bio Interface Center is gratefully acknowledged.

**Supporting Information Available.** Additional TEM images of a sample before and after in-situ current annealing, control experiment to study effects of carbon contamination, TEM images of two GNCs of width 6 and 1 nm, further analysis of data in Figure 4 (b). This material is available free of charge via the Internet at http://pubs.acs.org.

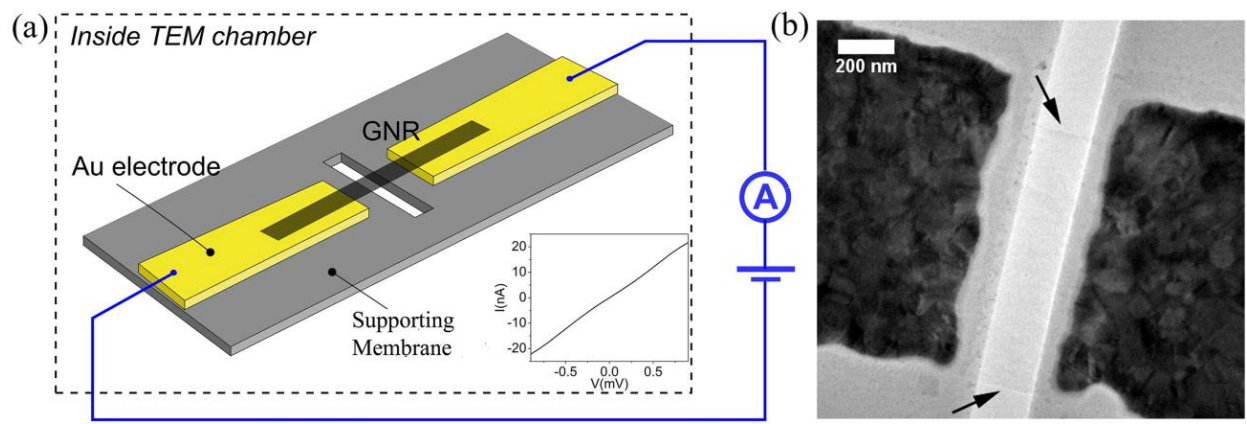

**FIGURE 1** Suspended graphene devices. (a) Sample schematic. Few layer graphene ribbon (3-10 layers thick) is suspended over a 1.4 $\mu$m×0.2 $\mu$m slit in a 100 nm thick silicon nitride (SiN) membrane (membrane size ~ 40 μm × 40 $\mu$m$^)$. Inset: Current-Voltage characteristic of an as-fabricated nanoribbon, acquired *in situ*. (b) TEM image of a suspended graphene nanoribbon. Arrows indicate the edges of the graphene.

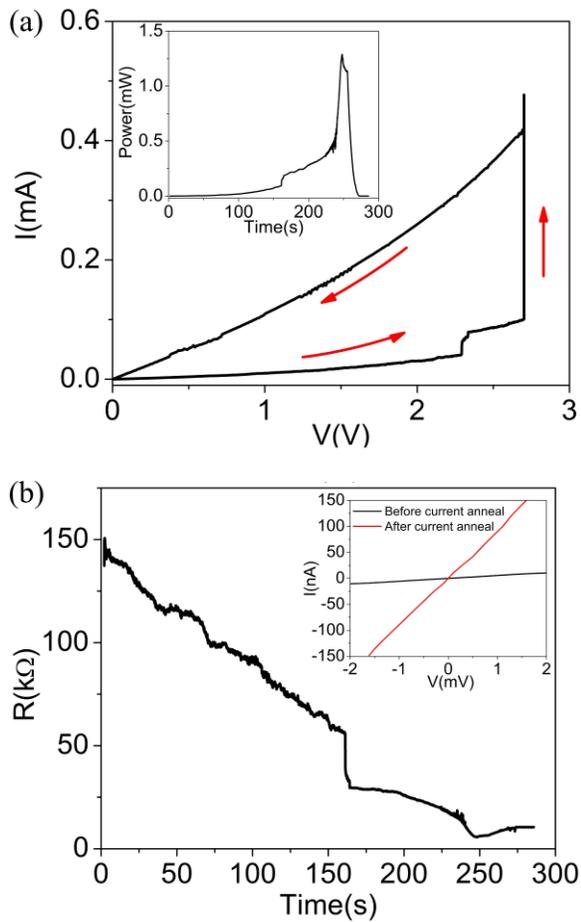

**FIGURE 2** Current annealing of graphene devices inside the TEM. (a) Current-Voltage (I-V) trace of a graphene ribbon device during current annealing; red arrows show current trace direction as the voltage is increased at a rate of 10 mV/s. Inset: corresponding plot of power vs. time. (b) Time evolution of the resistance (R=V/I) of graphene nanoribbon during current annealing. Inset: low bias I-V curve of the same device before and after current annealing.

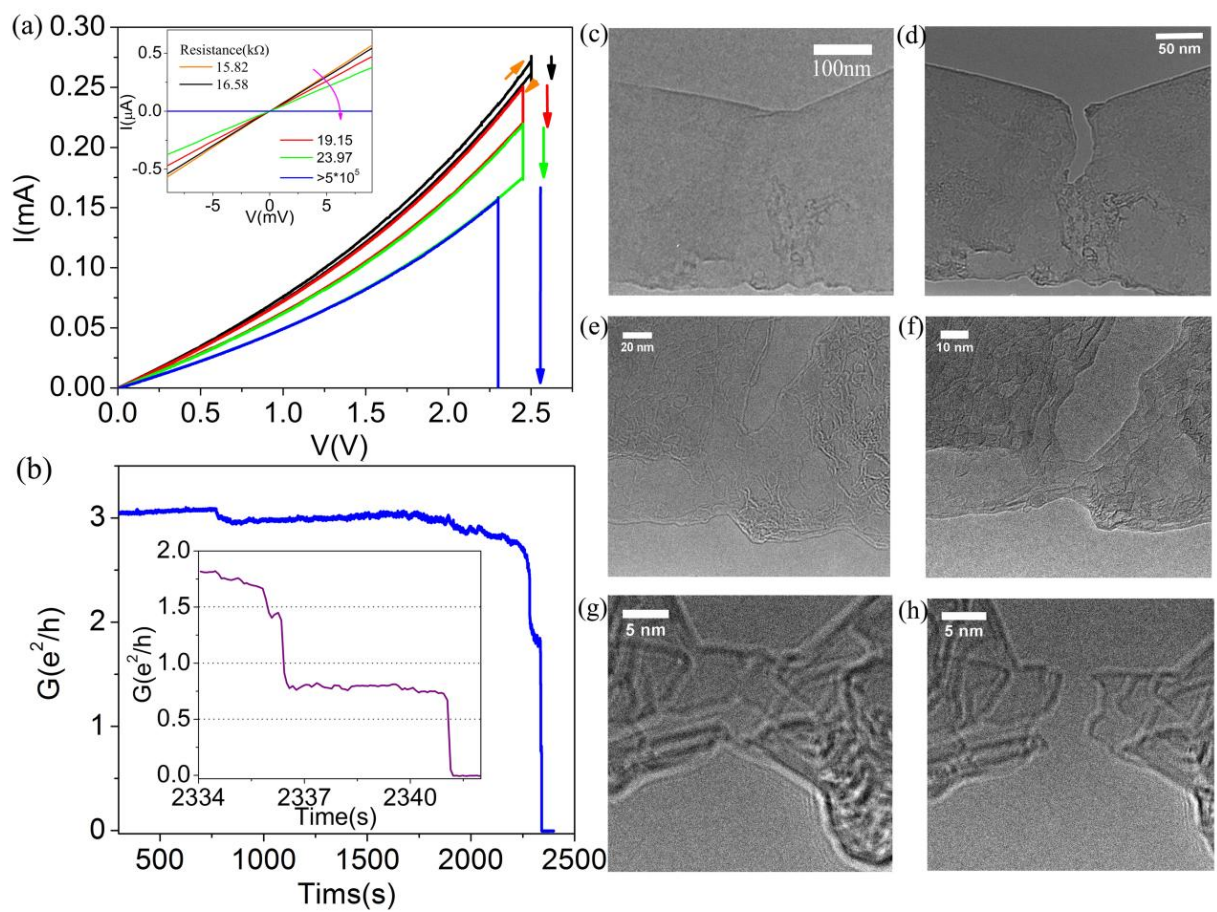

**FIGURE 3** TEM *in-situ* fabrication of graphene nanoconstriction. (a) Current-Voltage (I-V) traces of graphene nanoconstriction during fabrication using electron beam sculpting. Black, red, blue and green curves are I-V curves after four subsequent sculpting steps to reduce the constriction width, and vertical arrows indicate the associated conductance drops. Orange arrows on the black trace indicate direction of the voltage sweep. (Inset) Low bias I-V curves measured after each sculpting step. (b) Conductance vs. time plot of graphene constriction between the end of the last sculpting step ($w = 5$ nm) and spontaneous breaking of the constriction ($w = 0$ nm) under voltage bias, where the contact resistance has been subtracted. Inset shows conductance vs. time over the last eight seconds. (c) to (h) are TEM images for the same GNCs after each subsequent sculpting step from a width of ($w = 280$ nm) to fully broken ($w = 0$ nm). Note that (g) and (h) were taken at a slightly different focal point to enhance the contrast of the very thin GNCs.

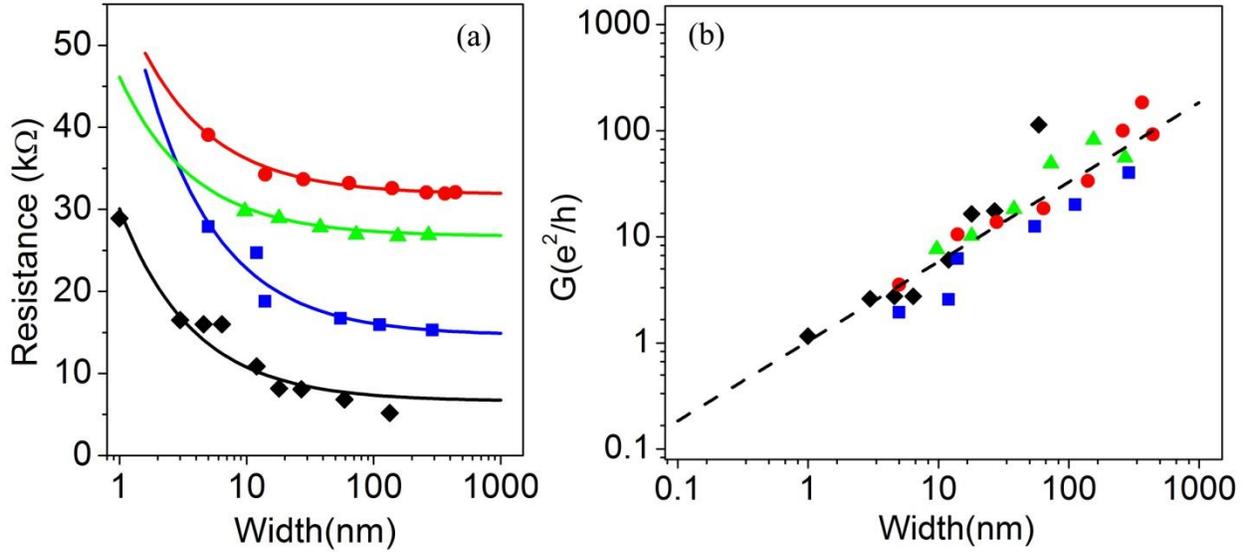

**Figure 4** Resistance and conductance as a function of device width. (a) Two-terminal GNC resistance as a function of width. Blue, green and red data are offset from each other by 10 kΩ for clarity; error bars are smaller than the symbol size. Fits are of the form $R_{TOT} = R_C + R_M \cdot w^{-\alpha}$ with $\alpha = 0.75$ and $w$, the GNC width measured in nm. Black fit: $R_C = 6.6$ kΩ, $R_M = 23$ kΩ; blue fit, $R_C = 4.6$ kΩ and $R_M = 46$ kΩ; green fit, $R_C = 6.4$ kΩ and $R_M = 19$ kΩ; red fit, $R_C = 1.8$ kΩ and $R_M = 24$ kΩ. (b) GNC conductance as a function of width with the contact resistance subtracted. Dashed line is a fit of the form $G \sim \sigma_0 \cdot w^{\alpha}$ with $\alpha = 0.75$, $w$ is the width of the GNC in nm, and $\sigma_0 = e^2/h$ ($R^2$ value of 0.86).

Table of Contents Graphic

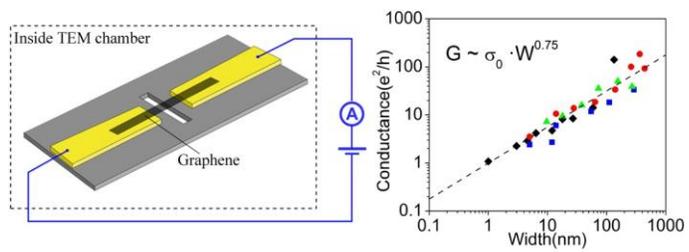